# Optically transparent and thermally efficient 2D MoS$_2$ heaters integrated with silicon microring resonators

*Dor Oz[1], Nathan Suleymanov[1], Boris Minkovich[1], Vladislav Kostianovskii[1], Liron Gantz[2], Dmitry Polyushkin[3], Thomas Mueller[3] and Ilya Goykhman[1]\**

[1] Micro Nanoelectronics Research Center, Department of Electrical and Computer Engineering, Technion, Haifa, Israel.

[2] NVIDIA Israel LTD, Yokneam, Israel.

[3] Institute of Photonics, Vienna University of Technology, Vienna, Austria.

KEYWORDS Integrated heaters, integrated photonics, 2D materials, MoS$_2$

ABSTRACT

Thermal tuning of the optical refractive index in the waveguides to control light phase accumulation is essential in photonic integrated systems and applications. In silicon photonics, microheaters are mainly realized by metal wires or highly doped silicon lines, placed at a safe distance (~1µm) from the waveguide to avoid considerable optical loss. However, this poses a significant limitation for heating efficiency because of the excessive free-carriers loss when a heater is brought closer to the optical path. In this work, we present a new concept of using optically transparent 2D semiconductors (e.g. $MoS_2$) for realizing highly efficient waveguide integrated heaters operating at telecom wavelengths. We demonstrate that a single-layer $MoS_2$ heater with negligible optical absorption in the infrared can be placed in close proximity (only 30nm) to the waveguide and show the best-reported heating efficiency of ~15 mW/FSR without sacrificing the optical insertion loss. The heater's response time is ~25 µs, limited by Au/1L-$MoS_2$ Schottky contact. Both the efficiency and response time can be further significantly improved by realizing 2D $MoS_2$ heaters with ohmic contacts. Our work shows clear advantages of employing 2D semiconductors for heaters applications and paves the way for developing novel energy-efficient, lossless 2D heaters for on-chip photonic integrated circuits.

Thermal tuning of the optical refractive index in waveguides is one of the basic operations performed in photonic integrated circuits (PIC) [1-2]. It is vital for a variety of applications including optical interconnect and telecommunication [3-5], smart and reconfigurable photonic networks [6-7], quantum photonics [8-9] and sensing [10]. Microheaters are widely employed in PIC to control phase accumulation of guided light via the thermo-optic effect, where temperature changes in the waveguide material induce the variation of the effective refractive index of the waveguide mode [11]. Today, thermo-optic heaters are widely utilized in PIC for thermal stabilization of photonic components [12], compensation for fabrication tolerances and design errors [13], spectral tuning and switching of integrated optical resonators [3,14], and execution of photonic and quantum operations based on optical interference and phase-shifting phenomena [8-9, 15-16]. In silicon photonics (SiPh) technology, microheaters are mostly implemented by metal wires [14, 17-18] or doped silicon lines [19-20] which are devised to dissipate electrical power to the heat due to Joule heating. However, these solutions pose considerable constraints and limitations on PIC design and performance [21]. Specifically, because of the free-carriers optical absorption in metals at telecom wavelengths (1.3-1.6 µm), metal heaters are separated from the waveguide mode by a sufficient gap (~1µm) [14, 16] to minimize optical insertion loss $\alpha_{IL}$ (Figure 1d). This separation leads to reduced heating efficiency $\eta_{th}$, increased power consumption (Figure 1c,d) and limited response time $t_r$ of metal heaters. The efficiency can be improved by conducting an electrical current directly through the waveguide [19-20]. However, this method requires doping of silicon causing an excess free-carriers optical loss, it cannot be implemented with insulating waveguide materials like silicon nitride (SiN) and it is not suitable for tunning active silicon photonic devices (e.g. PN-junction microring resonator). Eventually, to improve $\eta_{th}$ and $t_r$, one should target bringing a heater in direct contact with the waveguide without compromising $\alpha_{IL}$ (Figure 1c). Recently, there were several attempts in this direction using two-dimensional (2D) graphene heaters [22-24]. Own to the 2D nature of single-layer graphene (SLG), being an atomically thin semi-metal [25-26], the cross-section overlap between the waveguide mode and SLG is minimized providing a smaller $\alpha_{IL}$. The latter allows bringing the graphene heater closer (~240nm) to the waveguide, improving $\eta_{th}$ with optimized $\alpha_{IL}$ [23] - yet not at the lossless level. In particular, placing SLG in close proximity to the waveguide (<200nm) would result in a considerable insertion loss (Figure 1c).

On the other hand, atomically thin transition metal dichalcogenides (TMDs) with a chemical formula of $MX_2$ (M-transition metal atom, e.g. Mo, W, Nb; X-chalcogen atom, e.g. S, Se, Te) form a diverse class of 2D semiconducting materials [27]. 2D-TMDs are appealing for advanced electronic, optoelectronic, excitonic and valleytronic studies and applications [28-29]. Molybdenum disulfide ($MoS_2$) is one of the most studied 2D-TMDs [31]. Multilayer $MoS_2$ is an indirect bandgap semiconductor with an energy gap of ~1.3eV [30] (i.e. absorption edge at near-infrared wavelengths ~950nm) that increases to a direct bandgap of ~1.8-1.9eV (visible wavelengths ~650nm) [31] when the material is thinned down to a monolayer. Single layer MoS$_2$ (1L-MoS$_2$) is transparent at telecom wavelengths (e.g. photon energy of 0.8eV at 1550nm wavelengths) and can conduct substantial electrical current densities (~700µA/µm) [32], making it an excellent candidate for the development of optically lossless thermo-optic heaters for SiPh.

Here we report on the design, fabrication, and experimental characterization of the first optically transparent and efficient 1L-MoS2 microheater integrated with a silicon microring resonator (MRR) operating at 1550nm wavelength. We demonstrate that a 1L-MoS$_2$ heater can be placed in close proximity (~30nm) to silicon MRR with no penalty on insertion loss. The measured heating efficiency is ~0.1nm/mW, which is translated to only ~15mW of heating power needed to shift the MRR resonance for a full free spectral range (FSR). The latter is the best value reported in the literature. The heater's response time is ~25 µs, limited by Au/1L-MoS$_2$ Schottky contact. Both the efficiency and response time can be further significantly improved by realizing 2D MoS$_2$ heaters with ohmic contacts [33]. Our results pave the way for developing novel energy-efficient, optically transparent, and lossless 2D semiconductor heaters for on-chip photonic integrated circuits.

We first studied by numerical thermo-optical simulations the performance ($\eta_{th}$ and $\alpha_{IL}$) of different waveguide-integrated heaters using a finite-element commercial solver (Lumerical [34] MODE and HEAT). Figure 1c shows the simulated $\eta_{th}$ of a two-dimensional heat source (2µm width, 250µm length) dissipating an electrical power $P_{el} = 5mW$ into heat (i.e. power density of 10µW/µm$^2$ ) above a single-mode silicon waveguide (400nm width, 220nm height). The heater is separated from the waveguide by a SiO$_2$ layer of different thicknesses (Figure 1). We simulated a steady-state temperature rise $\Delta T$ in the waveguide for different heater-to-waveguide distances (gaps) and calculated the refractive index changes in silicon using $\Delta n_{Si} = dn/dT \cdot \Delta T$, where $dn/dT = 1.8 \cdot 10^{-4}$ is the thermo-optic coefficient in silicon [11,35]. Then, we extracted from simulations (MODE solver) the effective refractive index change $\Delta n_{eff}$ of the silicon waveguide mode produced by $\Delta n_{Si}$. For MRR integrated heater, the induced resonance wavelength shift $\Delta \lambda_{res}$ is given by $\Delta \lambda_{res} = \Delta n_{eff}/n_{eff} \cdot \lambda_{res}$, where $n_{eff} \sim 2.37$ is the simulated effective refractive index of the fundamental transverse-electric (TE) waveguide mode and $\lambda_{res} = 1550 nm$ is the MRR resonance wavelength. The heating efficiency is then given by $\eta_{th} = \Delta \lambda_{res}/P_{el}$. From Figure 1c we get that $\eta_{th}$ increases as the gap decrease, showing an improvement of ~30% for a gap thickness of 10nm compared to the 1μm distance typically employed for metal heaters [14, 16].

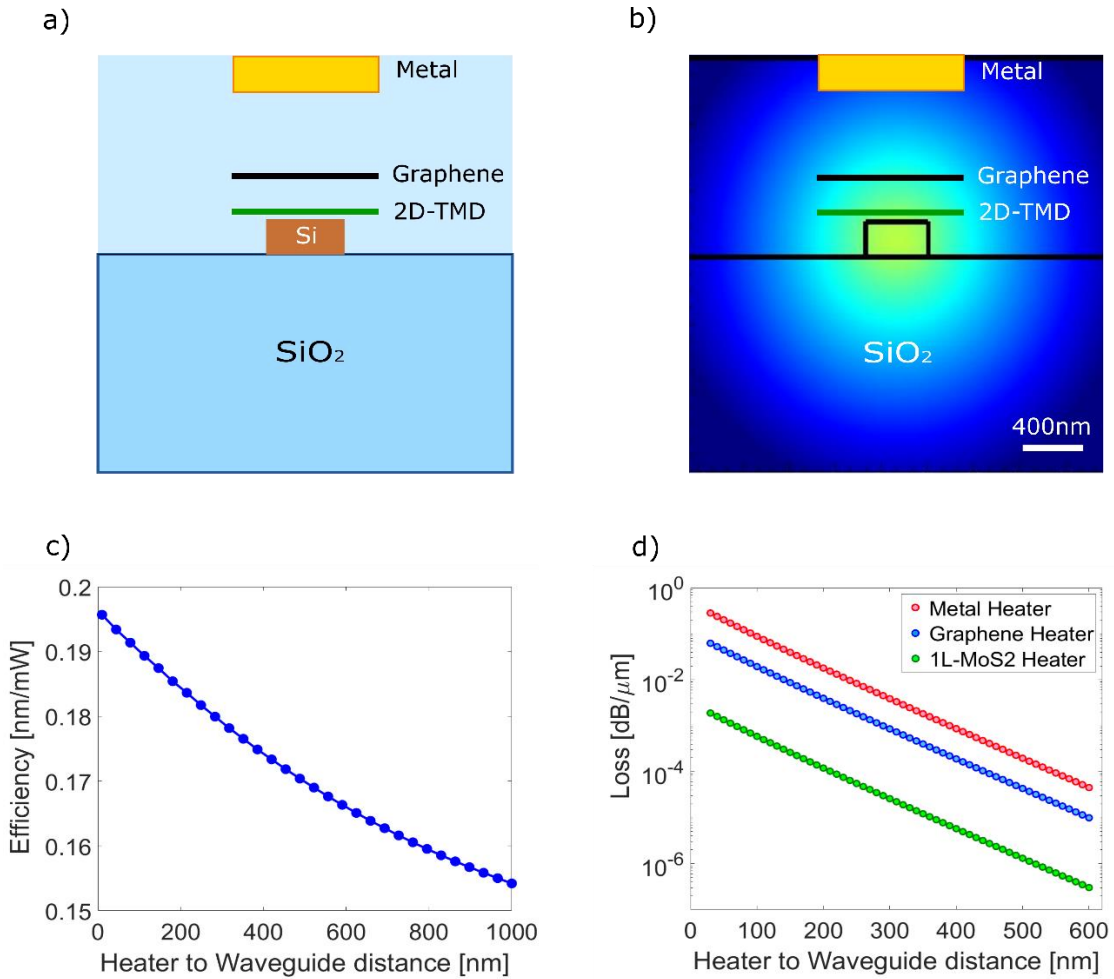

**Figure 1**. (a) Schematic comparison of separation distance (gap) from silicon waveguide to metal, graphene, and 2D-TMD heaters required to minimize the optical insertion loss. (b) Simulated optical intensity mode profile (logarithmic scale) of a single-mode silicon waveguide integrated with metal, graphene, and 2D-TMD heaters. The logarithmic scale is used to highlight the extension of the optical waveguide mode into a $SiO_2$ cladding and the evanescent interaction with metal, graphene, and 2D-TMD heaters. (c) Finite-element thermo-optical simulation of a steady-state temperature rise in a single-mode silicon waveguide as a function of heater-to-waveguide separation distance (gap). The simulated heating power density is 10 µW/µm$^2$ and the simulation results show increased waveguide temperature for smaller gap distances of a heater from the waveguide. (d) Finite-element optical simulation of optical insertion loss contributed by metal (tungsten), SLG and 1L-MoS$_2$ heaters as a function of heater distance (gap 30-500nm) from the waveguide. The simulated optically transparent 1L-MoS$_2$ heater shows a negligible ($<2 \cdot 10^{-3}$ dB/µm) insertion loss even when placed in direct contact with the waveguide.

Figure 1d shows the simulated $\alpha_{IL}$ for different heater materials (tungsten, SLG and 1L-MoS2) as a function of gap distance (30-500nm) from the waveguide. The optical properties of SLG were derived from its complex optical conductivity [26, 36] considering both the inter- and intraband absorption (i.e. no Pauli blocking). Specifically, we assumed the graphene Fermi level (doping) of $E_F = 0.2 eV$ and charge carriers mobility µ ~2000 cm$^2$/Vs (scattering rate $\Gamma = 13.5 \cdot 10^{12} \, s^{-1}$), corresponding to typical values of a chemical vapour deposition (CVD) grown polycrystalline SLG transferred on $SiO_2$. The optical properties of 1L-MoS$_2$ and tungsten were adopted from Ref. 37 and Ref. 38 respectively. The simulation results (Figure 1d) show that both the metal and SLG heaters induce considerable insertion loss for smaller (<200nm) gap values, while the simulated 1L-MoS2 heater shows a negligible $\alpha_{IL}$ (~$2 \cdot 10^{-3}$ dB/µm) even when it is in direct contact with the waveguide. Therefore, placing the MoS$_2$ heater in close proximity to the waveguide is expected to benefit from increased efficiency and lower power consumption due to better thermal coupling together with optical transparency and negligible loss.

The performance metrics of waveguide-integrated 1L-MoS$_2$ heaters were experimentally studied by using silicon MRR (Figure 2) and monitoring its transmission spectrum, $\alpha_{IL}$ and $t_r$ as the heat is dissipated in the microheater. The MRRs employed for this study were fabricated on a standard silicon-on-insulator (SOI) substrate with a 220nm silicon device layer on top of 2µm buried oxide (BOX). The ring resonators were defined by e-beam lithography followed by fluorine chemistry-based reactive ion etching (RIE), see Methods. To smooth the photonic chip surface before 1L-MoS$_2$ transfer we performed a spin-coating and thermal annealing of hydrogen silsesquioxane (HSQ) e-beam resist to realize a planarized surface with a shallow ~30nm thick SiO$_2$ layer on top of the silicon waveguide (Figure 2a,b). 1L-MoS$_2$ was grown by CVD on sapphire using solid precursors and then was transferred on a planarized silicon MRR using a semi-dry transfer method (see Methods). The 1L-MoS$_2$ heaters were patterned by additional e-beam lithography and RIE process steps, followed by electron-gun assisted metal evaporation of gold (Au) contacts. The scanning electron microscope (SEM) micrograph of the fabricated device is shown in Figure 2c,d.

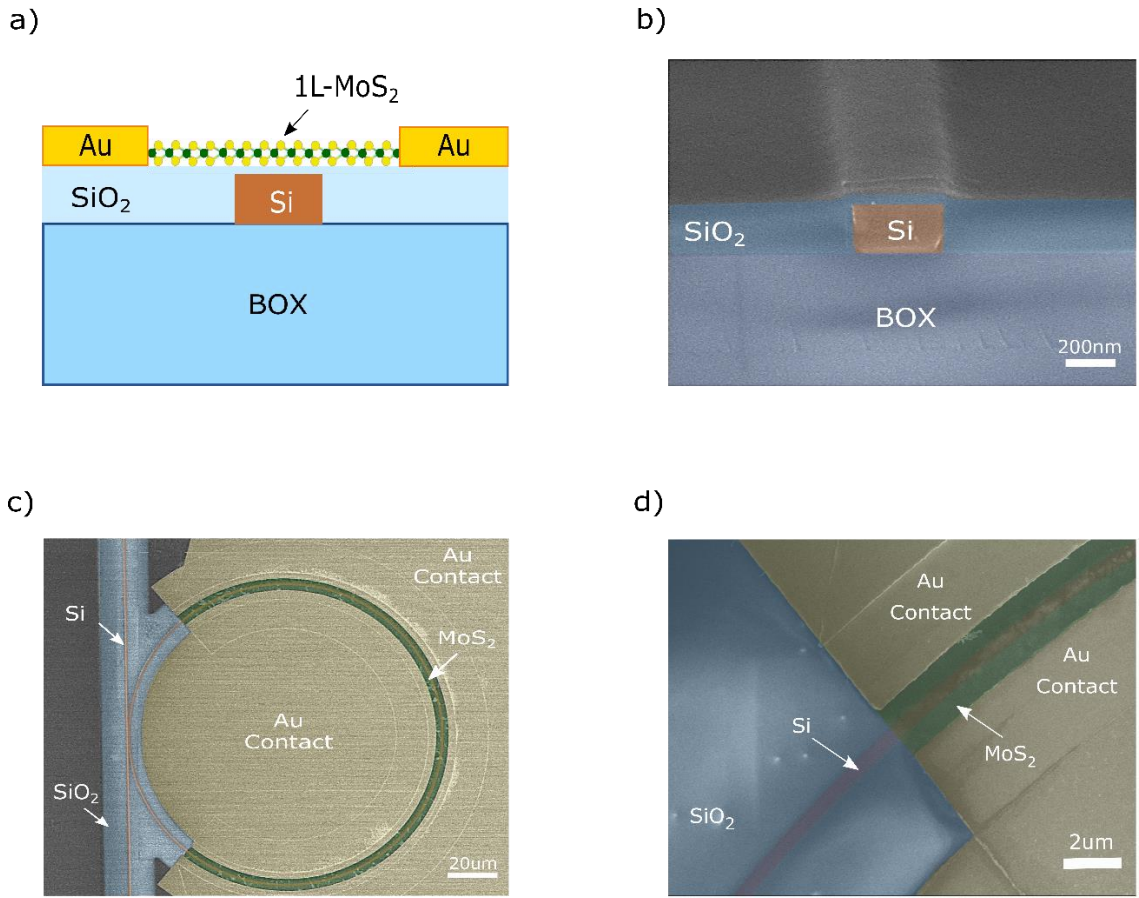

**Figure 2**. (a) Schematic cross-section of 1L-MoS$_2$ heater integrated on top of silicon waveguide. (b) Scanning electron micrograph of the planarized silicon waveguide (false color, cross-section view). The fabricated waveguide dimensions are 400nm in width and 220nm in height, supporting the fundamental TE-like waveguide mode. A planarized surface with a shallow ~30nm thick SiO$_2$ layer on top of the silicon waveguide (i.e. the 1L-MoS$_2$ heater to waveguide gap) is realized by an optimized spin-coating and thermal annealing of HSQ resist. (c) Scanning electron micrograph of the fabricated device (false color, top view). The 1L-MoS$_2$ heater length is ~283µm, covering ~3/4 of the silicon MRR perimeter, and its width is ~2µm as defined by the distance between source and drain Au contacts connected to the MoS$_2$ layer. (d) Scanning electron micrograph (false color) of 1L-MoS$_2$ heater on top of silicon MRR.

To monitor the quality and uniformity of the MoS$_2$ layer we performed Raman and photoluminescence (PL) spectroscopy characterizations (see Methods) of the as-grown material on sapphire and after the full device fabrication when MoS$_2$ heater is integrated on top of silicon MRR. Figure 3a (red curve) plots the Raman spectrum of CVD MoS$_2$ on sapphire for 532 nm excitation. The peak at ~386 cm$^{-1}$ corresponds to the in-plane ($E_{2g}^1$) mode [39-40] while that at ~406 cm$^{-1}$ is the out-of-plane ($A_{1g}$) mode [39-40] with full width at half-maximum FWHM ($E_{2g}^1$) = 3.1 cm$^{-1}$ and FWHM($A_{1g}$) = 3.8 cm$^{-1}$, respectively. The $E_{2g}^1$ mode softens, whereas the $A_{1g}$ stiffens with increasing layer thickness, [41-42] so that their frequency difference can be used to monitor the number of layers [41]. The measured peak position difference of ~19.5 cm$^{-1}$ indicates that a 1L-MoS$_2$ is used in our process [41]. The peak at ~417 cm$^{-1}$ (marked by an asterisk in Figure 3a) corresponds to the $A_{1g}$ mode of sapphire [43]. The minor (<1.5 cm$^{-1}$) red-shift of 1L-MoS$_2$ Raman spectra on the device is attributed to the difference in the dielectric permittivity between sapphire and SiO$_2$ (i.e. HSQ) layer, and the expected increase of n-type doping of 1L-MoS$_2$ after transfer on SiO$_2$ (i.e. top optical cladding of MRR) [44-45].

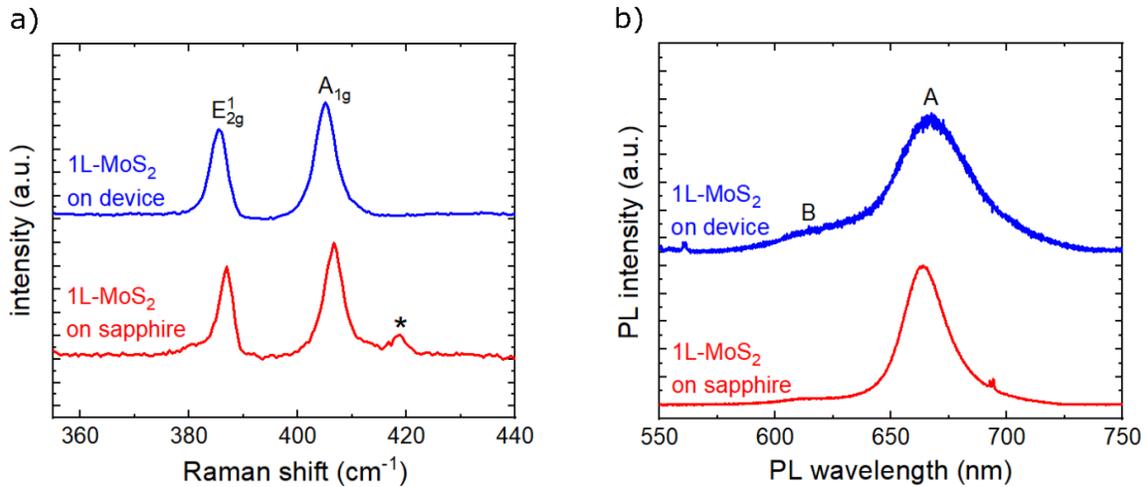

**Figure 3**. (a) Raman spectra at 532 nm of 1L-MoS$_2$ on sapphire and the device (i.e. HSQ optical cladding) after the full fabrication process. (b) PL spectrum at 532 nm (2.33 eV) of 1L-MoS$_2$ on sapphire and the device after the full fabrication process.

Another evidence for 1L-MoS$_2$ comes from the PL spectra (Figure 4b, red curve), showing characteristic A-exciton (~664 nm, ~1.87 eV) and B-exciton (~615 nm, 2.00 eV) peaks, a signature of band-to-band radiative recombination in 1L-MoS$_2$ [46]. The PL spectra on the device after the full fabrication process (Figure 4b, blue curve) reveal the broader and lower intensity A-exciton peak compared to the measurements of as-grown material on the sapphire (Figure 4b, red curve). This is attributed to enhanced exciton scattering and increased n-type doping of 1L-MoS$_2$ after the transfer on SiO$_2$. The observed decrease of A-exciton peak intensity is accompanied by enhanced charged exciton A$^-$ (trion) peak [47] as is evident from the PL spectra deconvolution (see Supplementary Information). The increased trion emission corresponds to a higher doping level of the 1L-MoS$_2$ layer after transfer. The minor (~1.5 nm) redshift of exciton peaks on the devices is ascribed to the difference in dielectric permittivities between sapphire and SiO$_2$ substrates [48].

The uniformity of 1L-MoS$_2$ after the full fabrication process was inspected by Raman and PL spectroscopy mapping of the heater. Figures 4b and 4c demonstrate a representative intensity map of Raman A$_{1g}$ mode and PL A-exciton peak across the device area (Figure 4a), showing a uniform signal around the waveguide (blue area). The waveguide profile is clearly identified in both maps revealing enhanced Raman and PL signatures (enhancement factor 2-3) when the signal is measured on top of a silicon waveguide covered with a ~30nm thick SiO$_2$ (i.e. HSQ) spacer. The effect can be explained by multiple reflections of the excitation wavelength from the underlying high refractive index silicon layer (i.e. waveguide, $n_{Si}$~4.15 @ 532 nm) acting as a back mirror and giving rise to enhanced interaction between the excitation signal and MoS$_2$ layer.

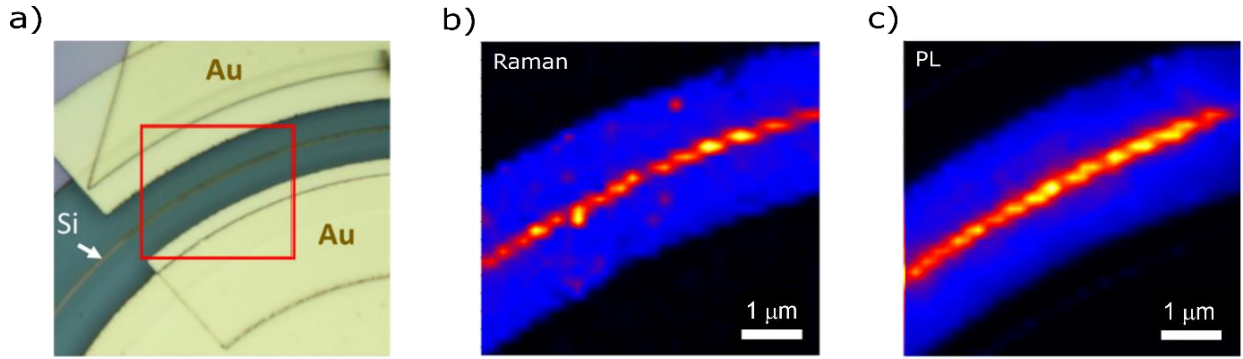

**Figure 4**. (a) Optical image of the mapped area (red square) of 1L-MoS$_2$ heater. (b) Intensity map of Raman A$_{1g}$ mode across the device area. (c) Intensity map of PL A-exciton peak across the device area.

The optical properties of ring resonators before and after MoS$_2$ integration were characterized by optical transmission measurements (Figure 5), where TE-polarized light from a tunable laser was coupled to MRR using a bus waveguide with a grating coupler and the transmitted optical signal was recorded at the output (through) port of the device. The transmission spectrum of bare MRR after planarization (Figure 5, blue line) shows well-defined resonance peaks with a quality factor $Q \sim 4000$, extinction ratio $ER > 10 dB$ and free-spectral range (FSR, spectral distance between two resonances) of $\Delta\lambda_{FSR} = 1.42 nm$. The transmission spectrum of MRR after MoS$_2$ heater fabrication is shown in Figure 5 (brown line). The observed spectral redshift is attributable to increased $n_{eff}$ of the ring waveguide due to the presence of MoS$_2$ (simulated $\Delta n_{eff} = 6 \cdot 10^{-4}$). The insertion loss of the MoS$_2$ integrated MRR remained unchanged (Figure 5) and the quality factor of the resonator showed small variations of <5%, demonstrating that the MoS$_2$ heater poses a negligible contribution to the loss. Specifically, after the full fabrication process, the integration of Au/1L-MoS$_2$/Au heater contributes ~$1.5 \cdot 10^{-3}$ dB/µm to the total loss of the device.

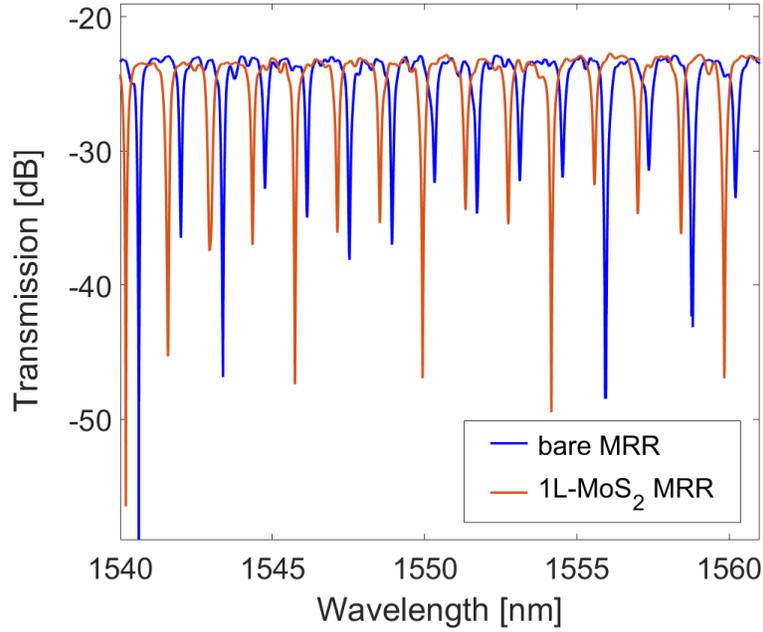

**Figure 5**. Optical transmission spectra of silicon microring resonator before (blue) and after (brown) MoS2 heater integration. The insertion loss of the device remained unchanged. The quality factor variation after the heater fabrication is <5%.

The electrical properties of the heater were studied by 2-terminal current-voltage (I-V) measurements and Kelvin probe force microscopy (KPFM) characterizations. The I-V plot (Figure 6a, blue line) shows a rectifying electrical characteristic suggesting a formation of a potential barrier (i.e. Schottky barrier) at the Au/1L-MoS$_2$ contact [49]. The latter is supported by KPFM measurements of electrostatic potential distribution across the heater indicating a built-in potential (i.e. a difference in work functions between the Au pad and MoS$_2$ channel) of ~130mV at the contact area, Figure 6c. We recorded the work function of gold to be higher than one of n-type semiconducting 1L-MoS$_2$ ($\phi_{Au} > \phi_{MoS_2}$), implying a Schottky contact band alignment at Au/1L-MoS$_2$ interface [50]. Therefore, the fabricated Au/1L-MoS$_2$/Au heater can be electrically described by two Schottky diodes connected back-to-back by a 1L-MoS$_2$ resistor. In this configuration, the obtained slight asymmetry in the I-V characteristics (Figure 6a) can be attributed to non-identical Schottky diodes having different areas (i.e. inner and outer contact of MRR) and anticipated variations in Schottky barrier height or ideality factor between the diodes [51-53].

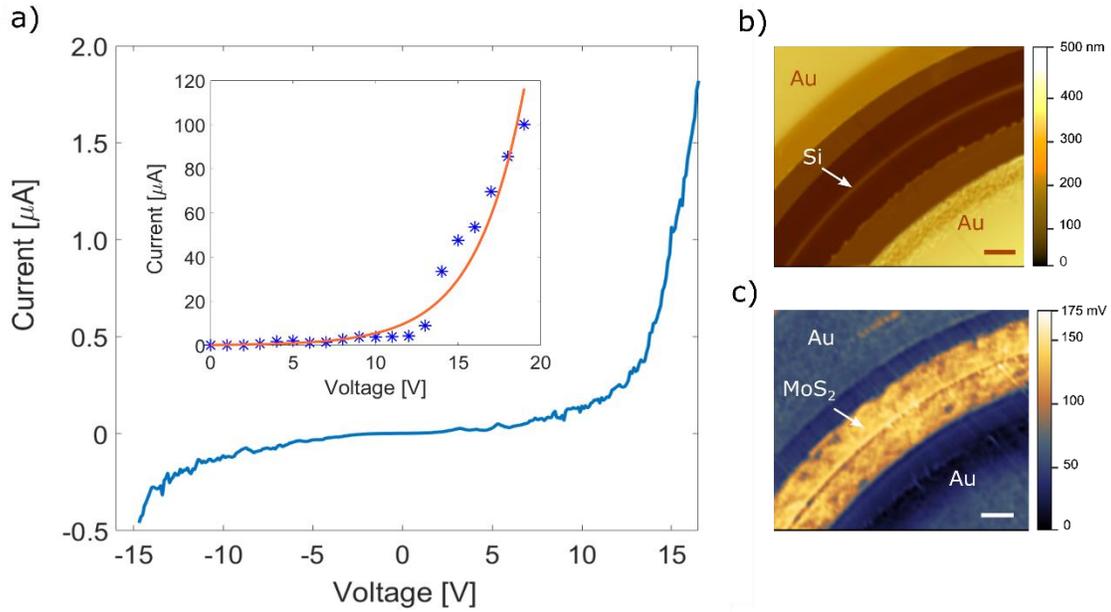

**Figure 6**. a) I-V characteristic of the MRR-integrated Au/1L-MoS$_2$/Au heater. Inset: The I-V characteristic of the heater when the current is measured with a long ~10s time delay after voltage bias application. b,c) Au/1L-MoS$_2$/Au heater area topography (b) and electrostatic potential distribution (c) measured by Kelvin probe force microscopy. The scale bar is 1μm.

The inset of Figure 6a shows the I-V characteristic of the heater when the current is measured with a long (~10s) time delay after voltage bias application. We achieved higher current levels when sufficient time was provided for current stabilization compared to values obtained using a short (~10ms) time-step, Figure 6a. The observed current increase could be attributed to the defect-mediated charge injection build-up across the Au/MoS$_2$ Schottky contact due to the charging/discharging of carriers through the interface traps [50, 54-56].

The performance of the microheater-waveguide configuration was studied by monitoring the optical transmission of the microring as heat is dissipated in the heater. Figure 7a shows the optical transmission spectra of MoS$_2$-integrated MRR for different heating powers. The resonances are redshifted (longer wavelengths) for increasing temperatures, corresponding to the positive thermo-optic coefficient of silicon [11,35]. The resonance wavelength shift as a function of heating power is presented in Figure 7b, where a linear fit to the data indicates the heating efficiency of ~0.1nm/mW, which is translated to ~15mW of electrical power required for a spectral shift of one FSR.

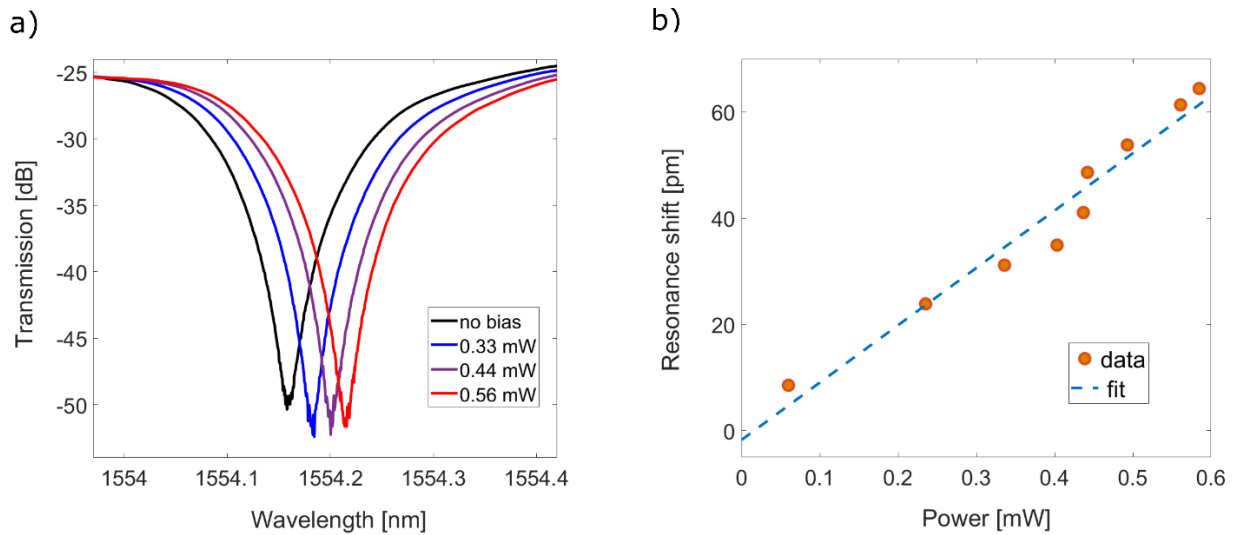

**Figure 7**. (a) Transmission spectrum of the ring resonance for different heater powers. (b) Resonance wavelength shift as a function of the heater power. The dashed line is a linear fit to the data with a slope of ~0.1 nm/mW (~15mw/FSR).

The power required to tune one FSR provides a geometry-independent comparison. To the best of our knowledge, the demonstrated efficiency $\eta_{th}$ ~15 mW/FSR is the best-reported value in the literature [22-24], showing the advantages of using optically transparent 2D semiconductors as integrated heaters with reduced power consumption when placed in close proximity (or direct contact) to the waveguide without sacrificing the optical insertion loss. The actual $\eta_{th}$ is even better considering that our heater covers only 3/4 of the ring circumference. It should be noted, that because of heat dissipation at the Au/MoS$_2$ Schottky contact, we experienced a low thermal breakdown power threshold of ~1mW, and therefore limited our experiments to current levels <100µA to preserve the contacts' integrity. Given the fact that a considerable fraction of the heating power is concentrated in the contact area and not uniformly distributed along the MoS$_2$ channel directly on top of the waveguide, we can assume that the heating efficiency can be significantly improved furthermore by realizing the 2D semiconducting heaters with ohmic contacts.

The response time of the thermo-optical tuning in the MoS$_2$/MRR device was characterized by measuring the rise $t_{rise}$ and fall $t_{fall}$ time of the temperature change in the resonator when the heater was biased with a square wave electrical signal (Figure 8, blue line). To operate within the linear regime of the resonator, we applied a DC bias of ~8V to the heater for setting the device operating point on the slope of the MRR resonance transmission function. To stay in the small-signal regime, a low AC excitation signal with 0.5V peak-to-peak amplitude at the frequency of 5 kHz was applied. The transmitted optical power was measured by a photoreceiver with a 17 kHz bandwidth connected to an oscilloscope. From the dynamic response (Figure 8, red line) we extracted $t_{rise}$ ~26 µs and $t_{fall}$ ~24 µs using a 10%-90% criterion with the average response time of the heater in the order of $t_r = (t_{rise} + t_{fall})/2 = 25$ µs.

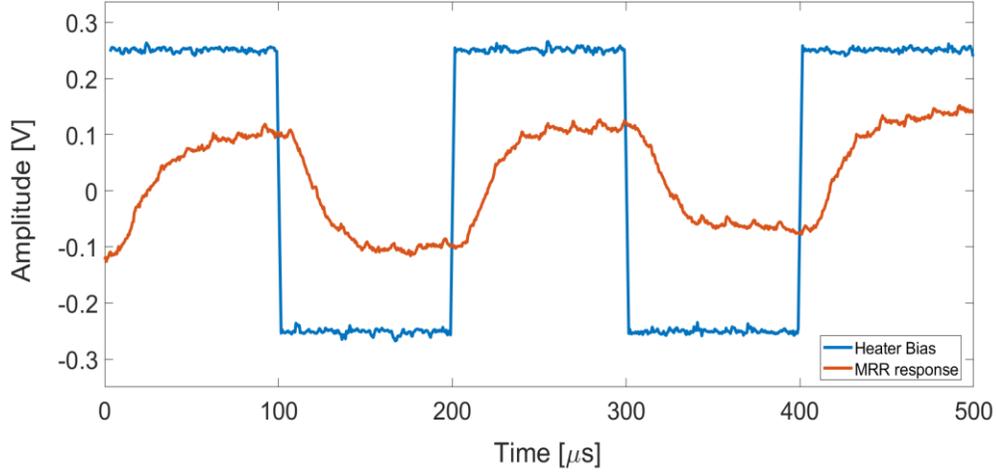

**Figure 8**. Time response of the integrated MoS$_2$/MRR device measured by detecting the transmitted light (red) during the application of a square wave small signal with 0.5V peak-to-peak amplitude at a frequency of 5kHz (blue). The extracted rise and fall times are 26 μs 24 μs respectively using the 10% to 90% criterion.

The obtained $t_r$ is comparable to the reported results on the conventional tungsten heaters on top of the waveguide without thermal insulation [18], however, it is slower compared to graphene heaters [23-24]. The response time in our device is limited by Au/1L-MoS$_2$ Schottky contacts, considering that the portion of the heat source (hot spot) is located ~1 μm off the waveguide, instead of being directly distributed on top of the waveguide if the power is mostly dissipated in the MoS2 channel and not at the contact area. Therefore, the response time can be considerably optimized by realizing metal/TMD ohmic contacts with low contact resistance.

In summary, we presented a new concept of using optically transparent 2D semiconductors (i.e. MoS$_2$) for realizing highly efficient waveguide integrated heaters operating at telecom wavelengths. The insignificant optical absorption of 1L-MoS$_2$ in the infrared allows placing the heater in close proximity to the waveguide (30nm) and achieving the best-reported heating efficiency of ~15 mW/FSR without sacrificing the optical insertion loss. The time response of the heater is ~25 µs and it is currently limited by the Au/1L-MoS$_2$ Schottky contact. Both the efficiency and time response can be further significantly improved by realizing 2D TMD heaters with ohmic contacts. Our work shows clear advantages in terms of power consumption and optical transparency for employing 2D semiconductors for heaters applications. Our results pave the way for developing novel energy-efficient, lossless 2D heaters for on-chip photonic integrated circuits.

**Methods**

As a substrate, we used a commercial SOI wafer with a 220 nm thick crystalline silicon device layer (p-type, 10-20 Ω·cm) on top of a 2 µm thick SiO$_2$ buried oxide layer. The photonic chip was patterned by electron beam lithography (EBL) using RAITH EBPG5200 writer and ZEP-520A e-beam resist. The pattern was reproduced in silicon using Inductively Coupled Plasma Reactive Ion Etching (ICP-RIE, Plasma-Therm) with C$_4$F$_8$/Ar/SF$_6$ gas mixture. A 250 nm thick HSQ layer (0.06 XR-1541, Dow Corning) was spin-coated and annealed at 200° C for 30 min to smooth the chip surface before the MoS$_2$ transfer.

The MoS$_2$ film was grown by CVD on sapphire according to the procedure described in Ref.[1], and then transferred onto the target wafer according to the modified procedure from Ref.[2]. The grown film is continuous over an area of >50 mm$^2$ with monolayer thickness.

As for the transfer, the chip with a CVD-grown MoS$_2$ was coated with polystyrene (PS) and immersed in 30% KOH solution for 1min and then in DI water for delamination. The PS/MoS$_2$ stack was picked up, washed, dried and laminated onto the photonic chip which was pretreated with O$_2$ plasma. After the transfer, the PS support layer was stripped with toluene exposing the 1L-MoS$_2$ for further processing. The MoS$_2$ heater was fabricated by an additional EBL patterning using a 300nm thick PMMA 950 A5 resist, following the RIE for shaping using $SF_6/O_2$ plasma and final metallization steps to realize 50 nm thick Au contacts to $MoS_2$ and Cr (5 nm)/Au (80 nm) contact pads to HSQ using e-gum evaporation and lift-off processes.

Raman and PL spectra were acquired using a Horiba LabRAM HR Evolution spectrometer at an excitation wavelength of 532 nm and a laser power of <0.5 mW to avoid sample heating or oxidation in air. An integration time of 2 s and 50 accumulations were used to enhance the signal-to-noise ratio for Raman measurements and 0.5 s integration time and 2 accumulations were used for PL measurements. The laser beam was focused onto the $MoS_2$ sample using a 100x objective lens with a NA of 0.9. The scattered light was collected and collimated by the same lens. The scattered signal was dispersed by a diffraction grating with 1800 grooves/mm (900 grooves/mm for mapping experiment) and then detected by a thermoelectrically cooled CCD (charge-coupled device) detector at -60 °C. All the Raman and PL spectra were recorded for the same integration time, laser power, and focus conditions.

The Kelvin Probe Force Microscopy characterizations were performed using Brucher-NanoIR 3 system. The topography and electrical potential measurements were acquired in the tapping and non-contact (~20nm from the surface) modes respectively, using Pt-coated silicon tips (Bruker PR-EX-KPFM) with a tip radius of ~ 30 nm and a resonance frequency of 67 kHz.

**Supporting Information**.

The following files are available free of charge.

Spectroscopic analysis of PL spectra (PDF)

**Corresponding Author**

Ilya Goykhman, email: ig@technion.ac.il

**Funding Sources**

We acknowledge funding from the EU Graphene Flagship, FLAG-Era and the Israel Innovation Authority (Grant No. 63350).

**Notes**

The authors declare no competing financial interest.

ACKNOWLEDGMENT

The devices were fabricated at the Micro-Nano Fabrication Unit at Technion.